\documentclass{article}
\usepackage[utf8]{inputenc}
\usepackage[T1]{fontenc}
\usepackage{hyperref}
\usepackage{url}
\usepackage{booktabs}
\usepackage{amsfonts}
\usepackage{nicefrac}
\usepackage{microtype}
\usepackage{xcolor}
\usepackage{amsmath,amsthm,amssymb}

\newtheorem{prop}{Proposition}

\newtheorem{defi}{Definition}[section]
\usepackage[
backend=biber,
style=alphabetic,
sorting=ynt
]{biblatex}
\title{Natural Language Mechanisms via Self-Resolution with Foundation Models}
\author{Nicolás Della Penna \\ nikete@grouplang.ai \thanks{I want to thank Tarun Chitra, Marc Graczyk and Jeff Strnad for helpful discussions.}}

\begin{document}
\maketitle

\begin{abstract}
Traditional mechanisms often constrain agent reports to simplified formats, potentially limiting expressible information. We propose Language Model Mechanisms (LMMs) that elicit natural language reports and leverage large language models (LLMs) for outcome selection and payoff assignment. We identify sufficient conditions for incentive-compatibility and efficiency: the LLM being a sufficiently good world model and a strong inter-agent information over-determination condition. We demonstrate LMMs can successfully aggregate information in scenarios where traditional mechanisms like prediction markets fail.
\end{abstract}

\section{Introduction}

Mechanism design, a cornerstone of economics and computer science, aims to create rules for social interactions that achieve desirable outcomes. Traditional mechanisms often constrain agent reports to simplified formats like trades or rank orderings, potentially limiting the information agents can express. This limitation can lead to suboptimal outcomes, especially when agents possess complex, high-dimensional private information.

We propose a novel class of mechanisms that leverage the power of large language models (LLMs) to overcome these limitations. Our main contributions are:

\begin{enumerate}
    \item We introduce Language Model Mechanisms (LMMs) that elicit agent reports in natural language, allowing for richer information exchange.
    \item We identify sufficient conditions for these mechanisms to be incentive-compatible and efficient, based on the LLM's capability as a world model and a strong inter-agent information over-determination condition.
    \item We demonstrate scenarios where LMMs can successfully aggregate information in signal structures where traditional mechanisms like prediction markets fail.
\end{enumerate}

Our approach represents a significant departure from conventional mechanism design, offering new possibilities for information aggregation and decision-making in complex environments.

\subsection{Motivating Examples}

To illustrate the potential of LMMs and the concept of information over-determination, consider the following scenarios:

\begin{enumerate}
    \item \textbf{Urban Development Planning:} A city plans a new high-density, walkable urban development. Potential residents describe their ideal living situations, daily routines, and community preferences. The over-determination condition is met because many participants will have overlapping preferences or complementary needs.
    
    \item \textbf{Collaborative Scientific Research:} In a large-scale scientific collaboration, researchers from various disciplines provide detailed reports on their findings, methodologies, and interpretations. The over-determination condition is satisfied because key scientific facts or methodological best practices will be known by multiple experts.
    
    \item \textbf{Crowd-Sourced Product Development:} A company uses an LMM to gather consumer insights for a new product line. Participants provide detailed descriptions of their needs, usage scenarios, and feature preferences. The over-determination condition is met because many consumers will have similar needs or use cases.
\end{enumerate}

These examples demonstrate how LMMs can leverage rich, natural language reports in settings where information is distributed across many agents with overlapping knowledge or experiences.

\subsection{Simple Example: 2 Variables and 6 Players}

Consider a scenario with two binary variables $X$ and $Y$, each taking values in $\{0,1\}$, and six players. The true state of the world, denoted by $Z$, is determined by the XOR of $X$ and $Y$:

\[ Z = X \oplus Y \]

Each player receives a noisy signal about one of the variables:

\begin{itemize}
    \item Players 1, 2, and 3 each receive an independent noisy signal about $X$
    \item Players 4, 5, and 6 each receive an independent noisy signal about $Y$
\end{itemize}

Let $s_i \in \{0,1\}$ be the signal received by player $i$. The signals are generated as follows:

\begin{align*}
    \Pr(s_i = X) &= 2/3 \quad \text{for } i \in \{1,2,3\} \\
    \Pr(s_i = Y) &= 2/3 \quad \text{for } i \in \{4,5,6\}
\end{align*}

In this setting, no individual player has enough information to determine the true state $Z$ with confidence greater than the prior. However, collectively, the players have sufficient information to determine $Z$ with high probability.

In a traditional prediction market for $Z$, if the prior probabilities for $Z$ are uniform, no individual player would have an incentive to trade, as their individual signal doesn't change the expected value of $Z$. 

In contrast, our language model mechanism can aggregate this distributed information effectively:

\begin{enumerate}
    \item Each player reports their signal to the mechanism in natural language.
    \item The LLM processes all six reports, recognizing that multiple consistent reports about X and Y provide strong evidence about their true values, and then computes $Z = X \oplus Y$.
    \item The LLM outputs its best estimate of the true state $Z$ based on all reports.
    \item The payment rule incentivizes truthful reporting.
\end{enumerate}

This example illustrates how our mechanism can successfully aggregate distributed information in scenarios where traditional mechanisms fail.

\section{Related Literature}

The literature on self-resolution, prediction markets, and peer prediction is vast. Recent work by Srinivasan et al. \cite{srinivasan2023self} shows there are truthful equilibria in a self-resolving prediction market under fairly standard assumptions for that literature. 

The capacity of LLMs to detect the miss reports when the information to do so is available, is related to the degree of  economic rationality of models, 
 \cite{raman2024steerassessingeconomicrationality} provide a framework to asses it.  There are, however, fundamental limits from pre-training approaches, that mean they must hallucinate if they are to be calibrated \cite{kalai2024calibratedlanguagemodelshallucinate}. 

\subsection{LLMs and Mechanism Design}

Several recent works have explored the intersection of LLMs and mechanism design:

\begin{itemize}
    \item \cite{hao2023reasoning} considers reasoning with language models as planning with a world model.
    \item Duetting et al. \cite{duetting2023mechanism} discuss mechanism design for large language models where the allocation and payment rules are constructed in a token-by-token manner over a set of LLMs.
    \item Dubey et al. \cite{dubey2024auctions} propose a factorized framework that contains an auction module and an LLM module.
    \item Rahaman et al. \cite{rahaman2024language} empirically study the reduction of the buyer's inspection paradox in information markets, showing cases where an LLM was able to reduce the asymmetry between buyer and seller in simulated scenarios.
\end{itemize}

The novelty in our work consists in using LLMs to elicit and aggregate rich information in natural language, with strong incentive guarantees under strong assumptions on the quality of the LLM and on the over-determination of the information across participating agents.

\section{Model}

Consider a set of agents $N = \{1, \ldots, n\}$. For each agent $i \in N$, let $S_i$ be a space of natural language signals. The joint signal space is denoted by $S = \times_{i \in N} S_i$. 

Let $s_i \in S_i$ denote the true signal observed by agent $i$, and let $s = (s_1, \ldots, s_n) \in S$ be the profile of true signals across all agents. We use $s_{-i}$ to denote the profile of true signals for all agents except $i$.

A \textit{language model mechanism} (LMM) consists of:

\begin{itemize}
    \item An \textit{outcome function} $f: S \to X$ that maps a profile of reported natural language signals to an outcome, using a large language model (LLM).
    \item A \textit{payment rule} $t = (t_1, \ldots, t_n)$, where $t_i: S \to \mathbb{R}$ specifies the payment to agent $i$ as a function of the report profile and the LLM's output.
\end{itemize}

The timing of the mechanism is as follows:

\begin{enumerate}
    \item Each agent $i$ observes their private signal $s_i$ and submits a report $r_i \in S_i$, where possibly $r_i \neq s_i$.
    \item The mechanism computes the outcome $x = f(r)$ and payments $t(r)$, where $r = (r_1, \ldots, r_n)$ is the profile of reports.
\end{enumerate}

In this model, we focus on the case where the information is separable from the preferences over the outcome. Agents with information only have preferences over their payments $t_i$, not over the selected outcome $x$ itself.

\subsection{Key Definitions}

\begin{defi}[$\delta$-Sufficient World Model]
An LLM is considered a $\delta$-sufficient world model if, for any profile of true signals $s \in S$, the outcome $x^*$ selected by the LLM satisfies:

\[ \mathbb{E}[W(x^*, s)] \geq (1-\delta) \mathbb{E}[W(x_{opt}, s)] \]

where $x_{opt} = \arg\max_{x \in X} W(x, s)$ is the optimal outcome, $W: X \times S \to \mathbb{R}$ is the welfare function, and $\delta \in [0, 1)$ is a small positive constant. The expectation is taken over any randomness in the LLM's output.
\end{defi}

\begin{defi}[Inter-agent Information Over-determination]
The information structure satisfies inter-agent information over-determination if, for any agent $i$ and any misreport $r_i \neq s_i$, either:
\begin{itemize}
    \item (Zero-shot setting) The LLM can detect with high probability that $r_i$ is inconsistent with respect to the true reports $s_{-i}$, without necessarily being able to reproduce $s_i$.
    \item (Observable outcomes setting) The expected forecasting error $\mathbb{E}[\varepsilon(r)]$ of the LLM increases when $r_i$ is substituted for $s_i$, i.e., $\mathbb{E}[\varepsilon(s)] < \mathbb{E}[\varepsilon(r_i, s_{-i})]$, where $\varepsilon: S \to \mathbb{R}_{\geq 0}$ is a forecasting error function and the expectation is taken over the distribution of signals for other agents.
\end{itemize}
\end{defi}

\subsection{Truthfulness and Efficiency}

The LMM has a \textit{truthful equilibrium} if for every agent $i \in N$, their true signal $s_i \in S_i$, all other agents' true signals $s_{-i} \in S_{-i}$, and any report $r_i \in S_i$:

$$\mathbb{E}[t_i(s_i, s_{-i})] \geq \mathbb{E}[t_i(r_i, s_{-i})]$$

That is, truthful reporting is a best response in expectation when others are truthful. The expectation is taken over any randomness in the LLM's output and the distribution of other agents' signals.

The truthful equilibrium is \textit{approximately efficient} if the outcome selected by the mechanism achieves expected welfare within a $(1-\delta)$ factor of the maximum expected welfare.

\begin{prop}
Under the following conditions, the language model mechanism (LMM) has a truthful and approximately efficient equilibrium:
\begin{enumerate}
    \item The LLM is a $\delta$-sufficient world model for some small $\delta > 0$.
    \item The information structure satisfies the inter-agent information over-determination condition.
\end{enumerate}
\end{prop}

\begin{proof}
We will prove this separately for the observable outcomes setting and the zero-shot setting.

\textbf{Observable Outcomes Setting:}

Let $\varepsilon: S \to \mathbb{R}_{\geq 0}$ be the forecasting error function of the LLM. Define the payment rule as:

$$t_i(r) = \alpha \cdot (K - \varepsilon(r))$$

where $\alpha > 0$ is a scaling factor and $K$ is a constant large enough to ensure non-negative payments.

For any agent $i$, true signal $s_i$, and potential misreport $r_i \neq s_i$, we have:

\begin{align*}
\mathbb{E}[t_i(s_i, s_{-i})] - \mathbb{E}[t_i(r_i, s_{-i})] &= \alpha \cdot (\mathbb{E}[\varepsilon(r_i, s_{-i})] - \mathbb{E}[\varepsilon(s_i, s_{-i})]) \\
&> 0
\end{align*}

The inequality follows from the inter-agent information over-determination condition in the observable outcomes setting.

\textbf{Zero-Shot Setting:}

Let $c: S \to [0,1]$ be a function representing the LLM's assessment of the consistency of a report profile. Define the payment rule as:

$$t_i(r) = \beta \cdot c(r)$$

where $\beta > 0$ is a scaling factor.

For any agent $i$, true signal $s_i$, and potential misreport $r_i \neq s_i$, we have:

\begin{align*}
\mathbb{E}[t_i(s_i, s_{-i})] - \mathbb{E}[t_i(r_i, s_{-i})] &= \beta \cdot (\mathbb{E}[c(s_i, s_{-i})] - \mathbb{E}[c(r_i, s_{-i})]) \\
&> 0
\end{align*}

The inequality follows from the inter-agent information over-determination condition in the zero-shot setting.

Thus, in both settings, truthful reporting is a best response in expectation when others are truthful, constituting a truthful equilibrium.

\textbf{Approximate Efficiency:} Given truthful reporting and the $\delta$-sufficient world model condition, the LLM selects an outcome $x^*$ such that $\mathbb{E}[W(x^*, s)] \geq (1-\delta) \mathbb{E}[W(x_{opt}, s)]$, ensuring approximate efficiency in expectation.
\end{proof}

\subsection{Information Structure}

The inter-agent information over-determination condition introduced in our model shares similarities with, but is distinct from, several classical concepts in information economics. In this section, we explore these relationships, focusing particularly on information monotonicity, which is a key concept in mechanism design.

\subsection{Information Monotonicity}

Information monotonicity is a condition often used in information economics to ensure that truthful reporting is optimal. In our context, we can define it as follows:

\begin{defi}[Information Monotonicity]
The information structure satisfies information monotonicity if for any agent $i$ and any misreport $r_i \neq s_i$,
\[ \mathbb{E}[\varepsilon(s_i, s_{-i})] \leq \mathbb{E}[\varepsilon(r_i, s_{-i})] \]
where $\varepsilon$ is the forecasting error function as defined in our model.
\end{defi}

This condition stipulates that truthful reporting leads to lower expected forecasting errors than misreporting. While our inter-agent information over-determination condition shares this intuition, it is in fact a stronger requirement, as we show in the following proposition:

\begin{prop}
The inter-agent information over-determination condition (in the observable outcomes setting) implies information monotonicity, but the converse is not true.
\end{prop}

\begin{proof}
First, we show that inter-agent information over-determination implies information monotonicity:

Let $i$ be any agent and $r_i \neq s_i$ be any misreport. 
By the inter-agent information over-determination condition, we have:
\[ \mathbb{E}[\varepsilon(s)] < \mathbb{E}[\varepsilon(r_i, s_{-i})] \]

Since $s = (s_i, s_{-i})$, this is equivalent to:
\[ \mathbb{E}[\varepsilon(s_i, s_{-i})] < \mathbb{E}[\varepsilon(r_i, s_{-i})] \]

This strict inequality clearly implies the non-strict inequality required by information monotonicity:
\[ \mathbb{E}[\varepsilon(s_i, s_{-i})] \leq \mathbb{E}[\varepsilon(r_i, s_{-i})] \]

To show that information monotonicity does not imply inter-agent information over-determination, we provide a counterexample:

Consider a setting with two agents, where $\varepsilon$ can take only two values: 0 or 1. 
Let the joint distribution of signals be such that:

$\mathbb{E}[\varepsilon(s_1, s_2)] = 0.5$
$\mathbb{E}[\varepsilon(r_1, s_2)] = 0.5$ for any $r_1 \neq s_1$
$\mathbb{E}[\varepsilon(s_1, r_2)] = 0.5$ for any $r_2 \neq s_2$

This setting satisfies information monotonicity, as for any $i$ and $r_i \neq s_i$:
$\mathbb{E}[\varepsilon(s_i, s_{-i})] = 0.5 \leq 0.5 = \mathbb{E}[\varepsilon(r_i, s_{-i})]$

However, it does not satisfy inter-agent information over-determination, because the strict inequality $\mathbb{E}[\varepsilon(s)] < \mathbb{E}[\varepsilon(r_i, s_{-i})]$ does not hold.
\end{proof}

\subsection{Other Related Concepts}

While our focus has been on information monotonicity, the inter-agent information over-determination condition also relates to other concepts in information economics:

1. \textbf{Single-Crossing Condition}: In mechanism design, the single-crossing condition often ensures that an agent's preferences satisfy a certain monotonicity property. Our condition similarly ensures a form of monotonicity, but with respect to the quality of information rather than preferences.

2. \textbf{Supermodularity}: In some information structures, the marginal value of one agent's information increases with the quality of other agents' information. While our condition doesn't directly imply this property, it does capture a related idea of information complementarity across agents.

\section{Discussion}
The key insight that language models are world models can be leveraged to make their use as allocation functions incentive compatible by linking the payoff of agents to their performance as scored by the model using the other reports. In practice, this allows for the exploration of a novel and expressive set of mechanisms, with very different assumptions than previously proposed mechanisms.
\subsection{Practical considerations}
The prompt template that the reports are inserted into effectively acts as contract. It is used when instantiating the prompt that is given to the LLM.
One practical approach is to use an intermediary representation that would be the direct output of the LLM, such as code that can be checked for syntactic correctness and tested. This code can then further be executed to generate the outcome and the payment.
In general, when we refer to the LLM we include pre and post processing pipelines that populate its prompt.

\subsection{Limitations \& Application Scope}
The conditions identified for sufficiency are very strong. In particular the information structure needs to have enough redundancy such that it is not possible to go undetected when making directionally correct reports that have a slight bias. This implies that any piece of information must be had by a sufficiently high number of agents.
There are potential application domains for institutions designed around such mechanisms, that the sufficient conditions might be met, namely:
The underlying information structure has a high degree of redundancy across agents. Some examples include:
\begin{itemize}
\item LLM agents that simulate individuals' preferences for who to have or not as neighbors.
\item LLM agents representing potential buyers on a new high density walkable urban development in a previously unpopulated area. Being able to coordinate who is neighbors with whom is hard without LLMs because most potential buyers do not have the information nor time to evaluate how good of a match they would be to all hypothetical potential neighbors. Having the LLM evaluate the compatibility of the reports (which might include both the potential buyers' stated preferences and their observable characteristics) could enable better coordination.
\item The miss-report detection might then be possible by using the reports of those who do know the individual about them that their LLM generates, and using this to evaluate the self-reported LLM simulator.
\end{itemize}
The linguistic report can be thought of in several ways:
\begin{itemize}
\item Most concretely: As a string reporting preferences
\item As a string that is used to prompt or fine-tune a given standard model
\item As a set of weights for a given standardized model architecture
\item Most generally: a pipeline and the weights of the LLM(s) used in it representing the user
\end{itemize}
An interesting direction for future work is in the most general model, if the prompts being used in the pipeline are themselves the subject of optimization (as in DSPy).
In the setting with feedback, to enable self-resolution we need the model to be able to evaluate the truthfulness of reports.

\subsection{Future work}
Key empirical challenges include understanding how and to what limits empirical measurements can determine if the conditions for truthfulness and efficiency are met, and how and to what extent can we measure the agents' information over-determation.
A key theoretical question is how much the agent information substitutability conditions can be relaxed. A natural and interesting generalization of the information structure is when the signal that the world model needs is intermixed with the private information of those with preferences over the outcome. In other words, in the setting where agents with information have preferences not only over their payments $t_i$ but also over the selected outcome $x$ itself. Under the truthful equilibrium with the assumption that deviations are zero-shot detectable, agents falsifying their reports to manipulate the allocation $x$ would be detectable. If these agents are punishable in the payments they could be made out of the equilibrium. In the observable outcomes settings, it would be valuable to understand how the error of the language model in forecasting relates to the degree to which agents can miss-report undetectably.

\printbibliography
\end{document}